\documentclass[superscriptaddress]{revtex4-1}

\pdfoutput=1

\usepackage{graphicx}
\usepackage{amssymb,amsfonts,amsmath}
\usepackage[usenames,dvipsnames]{color}
\usepackage{natbib}

\newcommand{\diffd}{\mathrm{d}} 
\newcommand{\conj}[1]{\overline{#1}} 
\newcommand{\real}[1]{\mathrm{Re} \left[ #1 \right]} 
\newcommand{\eee}[1]{\mathrm{e}^{ #1 }} 
\newcommand{\ii}{\mathrm{i}} 

\newcommand{\fig}[2]{Fig.~\ref{#1}\textit{#2}}
\newcommand{\figs}[2]{Figs.~\ref{#1}\textit{#2}}

\begin{document}

\title{Learning to school in the presence of hydrodynamic interactions}
\author{Mattia Gazzola}
\affiliation{School of Engineering and Applied Sciences, Harvard University, Cambridge, MA 02138, USA}
\author{Andrew A. Tchieu}
\affiliation{SPACEX, Hawthorne, CA 90250, USA}
\author{Dmitry Alexeev}
\affiliation{Department of Computational Science, ETH Z\"urich, CH-8092, Switzerland}
\author{Alexia de Brauer}
\affiliation{Institut de Math\'ematiques de Bordeaux, UMR CNRS 5251, Universit\'e Bordeaux 1 et Inria Bordeaux - Sud Ouest 351 cours de la Libration 33405 Talence, France}
\author{Petros Koumoutsakos}
\email{petros@ethz.ch}
\affiliation{Department of Computational Science, ETH Z\"urich, CH-8092, Switzerland}

\begin{abstract}
Schooling, an archetype of collective behavior, emerges from the interactions of fish responding to visual and other informative cues mediated by their aqueous environment. In this context, a fundamental and largely unexplored question concerns the role of hydrodynamics. Here, we investigate schooling by modeling swimmers as vortex dipoles whose interactions are governed by the Biot-Savart law. When we enhance these dipoles with behavioral rules from classical agent based models we find that  they do not lead robustly to schooling due to flow mediated interactions. In turn, we present dipole swimmers equipped with adaptive decision-making that learn, through a reinforcement learning algorithm,  to adjust their gaits  in response to non-linearly varying hydrodynamic loads. The dipoles maintain their relative position within a formation by adapting their strength and school in a variety of prescribed geometrical arrangements. Furthermore, we identify schooling patterns that minimize the individual and the collective swimming effort, through an evolutionary optimization. The present work suggests that the adaptive response of individual swimmers to flow-mediated interactions is critical in fish schooling.
\end{abstract}

\maketitle

\section{Introduction}
Schooling, encountered in over ten thousand species \citep{Shaw:1978}, is believed to provide several advantages to fish \citep{Partridge:1982} including protection and defense against predators \citep{Shaw:1978, Major:1978, Landeau:1986}, enhanced  foraging \citep{Pitcher:1982} and mating success \citep{Barnes:1988}. It is also plausible that fish benefit from increased hydrodynamic efficiency \citep{Weihs:1973}. Understanding the governing mechanisms in fish schooling and exploiting them for rational engineering designs \citep{Whittlesey:2010} requires that we elucidate the interplay between social and hydrodynamic interactions among swimmers.

While such distinctions may be difficult to {investigate} in experimental or natural settings, the detailed information  that can be obtained via simulations {are} invaluable. At the same time while schooling can be readily observed in natural and experimental settings, in simulations it is essential to equip the individuals with an appropriate behavioral model to achieve such group dynamics. Agent based models \citep{Reynolds:1987} {that lead to schooling or flocking rely on local interaction rules handcrafted \textit{a priori} based on empirical arguments and experimental observations \citep{Couzin:2005, Couzin:2002, Viscido:2005, Hubbard:2004}. These models} have been a key tool in helping to understand the influence of social {traits} in the emergence of schooling patterns \citep{Aoki:1982, Huth:1992, Niwa:1994,Vicsek:1995}. However, they do not explicitly account for the flow environment. We consider this a limitation especially in the case of large, tightly packed fish assemblies. In fact, a natural swimmer which wishes to adapt its speed and orientation to satisfy local interaction rules (e.g. move with the average velocity of its neighbors) needs to translate this into specific body gaits. These actions perturb the flow field, which in turn affects the dynamics of the neighbors.

It is also important to distinguish between self-propelled swimmers and swimmers that are towed with a specified velocity through the flow field \citep{Jiang:2002}, as it is usually implied in agent based models. In the case of a towed swimmer, {if hydrodynamics is included it only affects} the energy expenditure for towing the swimmer with the specified velocity, \textcolor{black}{while it does not influence its} dynamics nor its trajectory. A self-propelled swimmer instead has to adjust its gait to compensate for {non-linearly varying} hydrodynamic loads to propel itself in a desired direction.  As fish rely on self-propulsion, it is essential to capture this trait altogether with the long range fluid coupling. To the best of our knowledge, such hydrodynamic interactions have not been included in agent based models of swimming. Hence, fundamental questions on  how fish respond to each other's wakes and to what extent is schooling the result of their synthesized vortex field or their social traits, remain largely unanswered. 

Swimmers influence their flow environment which in turn affects the dynamics of the individual at all scales. In the Stokes flow regime, it has been noted that the collective motion of microorganisms induces flow coupling that leads to transitions from ordered to disordered patterns \citep{Hatwalne:2004,Brady:1988,Brady:1988a,AditiSimha:2002}. Both in the inviscid limit and at finite Reynolds numbers, recent works have demonstrated that specific body motions can propel initially stationary neighbors \citep{Tchieu:2010a, Gazzola:2012a}, while models of rotating discs at finite Reynolds numbers have been shown to lead to the emergence of patterns \citep{Goto:2015}. Experimental observations indicate that some fish species arrange themselves in diagonal formations \citep{Cullen:1965, Partridge:1980a} and  it has been suggested \citep{Breder:1965,Weihs:1973, Tsang:2013} that fish in diamond configurations can exploit the vorticity created by their neighbors to decrease their energy expenditure. This hypothesis relies on stable, periodic fish arrangements, prescribed gait and unperturbed or minimally perturbed flow conditions. At intermediate and large Reynolds numbers the flow field synthesized by the vorticity shed by multiple swimmers \citep{Abrahams:1987,Liao:2003,Weihs:2004,Ristroph:2008} is noisy, and varying loads are induced on the swimmers depending on their relative location \citep{Gazzola:2011a}. How can swimmers overcome this noisy environment to achieve specific behavioral or physical goals? To what extent flow-mediated interactions affect decision making and group behavior?  

In this article we  investigate the effect of flow-mediated interactions on the internal structure and global shape of schools composed by hundreds of model dipole swimmers. This work is inspired by the concept of {\it Vortobots} \citep{Park:2005}. The Vortobots were envisioned as simplified rotating bodies (vortices) that move in swarms by controlled hydrodynamic interactions. Here, following Tchieu et al. \citep{Tchieu:2012}, swimmers are modeled as self-propelled, finite width dipoles capable of accelerating, decelerating and turning. We show that the use of non-adaptive \textit{a priori} defined local interaction rules does not robustly allow swimmers to maintain finite size schooling formations, causing them to diverge from one another or to collide. In turn we show that swimmers can learn, through a reinforcement learning (RL) algorithm \citep{Sutton:1998}, to dynamically adjust their swimming actions in response to flow-mediated interactions so as to swim in arbitrary finite size schooling arrangements. Furthermore, we identify schooling arrangements that minimize collective swimming effort, via an evolutionary optimization technique. Finally, the relative effort of swimmers distributed within an optimal school is investigated. 

Our study highlights the importance of accounting for the hydrodynamic environment in collective dynamics, and outline a rigorous methodology for identifying optimal adaptive action policies so as to respond to flow-mediated interactions.

\section{Learning optimal behavior in a fluid-mediated environment}
\label{sec:methods}
We examine the collective behavior of model dipole swimmers. In order to control their velocity and bearing, the swimmers can adapt their dipole strengths and as such they affect the environment and, in turn, the dynamics of all other swimmers. In contrast to classical agent based models \citep{Aoki:1982, Huth:1992, Niwa:1994,Vicsek:1995}, besides including hydrodynamics, we do not specify \textit{a priori} local interaction rules. Instead, these are automatically identified by a reinforcement learning algorithm.

\begin{figure}
	\begin{center}
		\includegraphics{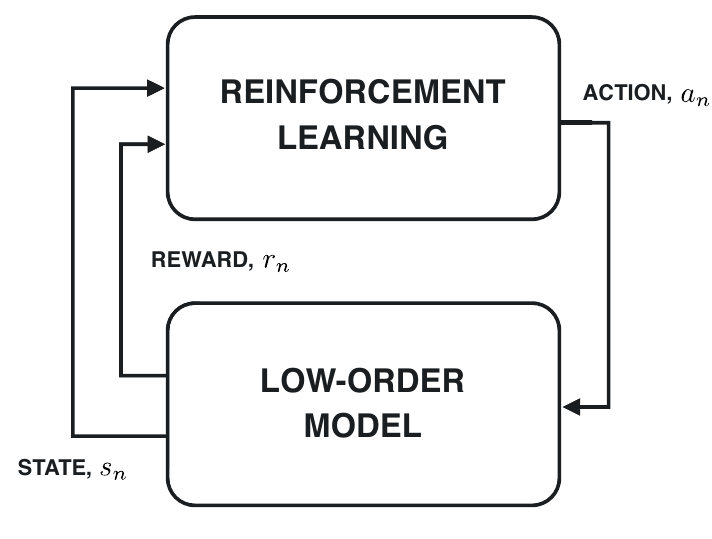}		
		\caption{Schematic of reinforcement learning coupled with a low-order model for swimmers in an inviscid flow. The goal of the $n$th agent is encoded into the numerical reward $r_{n}$ and the agent learns, through a trial and error process called reinforcement learning \citep{Sutton:1998}, how to map states $s_{n}$ into actions $a_{n}$ to maximize the long term reward (Section \ref{sec:rl}). The system is simulated via a low-order model that takes into account swimmer-swimmer dynamics mediated by an inviscid fluid medium \citep{Tchieu:2012}.}
		\label{fig:approach}
	\end{center}
\end{figure}

The dynamics of a system of $N$ swimmers {}{immersed in an inviscid flow} is represented as a low-order model denoted as `finite width dipole' \citep{Tchieu:2012}. The goal of the swimmers is to swim coherently in a prescribed formation, avoiding collisions or dispersion. This is a necessary preliminary step to allow for the hydrodynamic characterization of different swimming formations in terms of energetic expenditure (Section \ref{sec:optimalSchools}). Given the swimmers repertoire of possible actions and sensorial representation of the environment (denoted as states), reinforcement learning \citep{Sutton:1998} allows them, through trial and error, to discover an optimal behavioral policy (i.e. a mapping between states and actions) to maintain their relative positions within the school. Each loop in \fig{fig:approach}{} represents a single learning instance where all agents use their learned policy to select an action which alters their state through the modeled dynamics. The reward associated with the new state aids the agents in improving their policy, which eventually converges to an optimal policy.

\subsection{A finite width dipole model for hydrodynamically interacting swimmers}
\label{sec:model}
\begin{figure}
	\begin{center}
		\includegraphics[width=\textwidth]{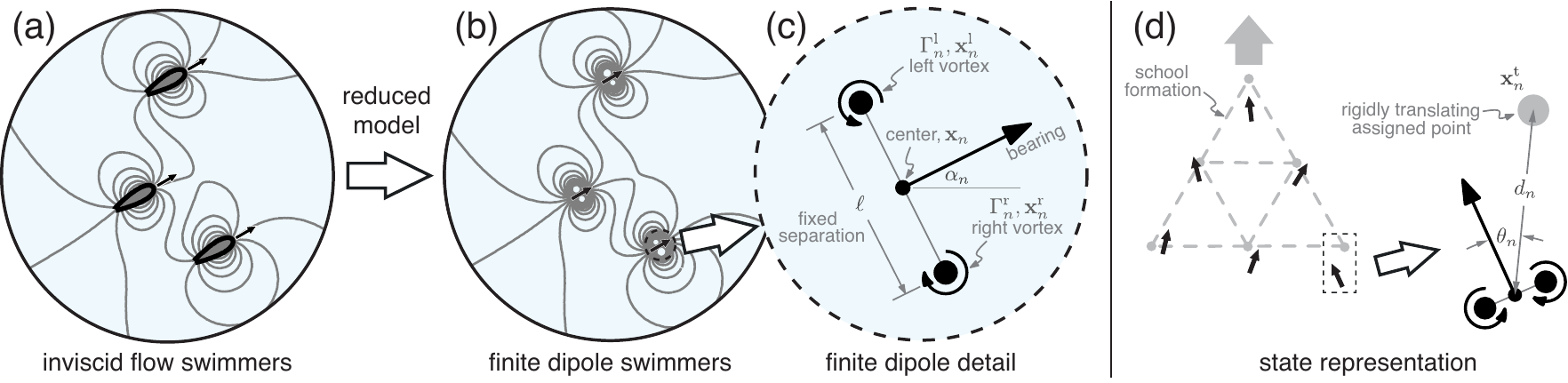}		
		\caption{Streamlines of (a) inviscid swimmers translating in a potential flow and (b) the finite dipole approximation of the swimmers \citep{Tchieu:2012}. Given in (c) is the detailed view of a single finite dipole. (d) illustrates the state representation of an individual dipole swimmer attempting to follow a translating lattice point (Section \ref{sec:rl}).}
		\label{fig:model}
	\end{center}
\end{figure}

The flow field generated by individual natural swimmers possesses a complex signature that is greatly affected by their gait, morphology and size as well as by viscous and three-dimensional effects. The characterization of group dynamics of hundreds of swimming bodies that resolves this level of detail is to date computationally beyond reach. Therefore, we study swimmers modeled as finite self-propelling dipoles (\fig{fig:model}{a--b}) immersed in an inviscid, unbounded and incompressible flow \citep{Tchieu:2012}. This model reflects the fact that the far field associated with a self-propelled undulating body is dipolar, to leading order \citep{Wolfgang:1999}. The finite dipole model represents a drastic idealization of a swimmer since it abstracts from morphological and kinematic traits, it is massless and therefore disregards the inertia of a solid body and does not account for three-dimensional and viscous effects, such as separation and vortex shedding. Nevertheless, it does capture at first order the flow coupling among self-propelled bodies sufficiently spaced apart (more than one body length as estimated in \cite{Tchieu:2012}), and it is computationally effective. Bearing in mind its  limitations, the dipole model renders itself instrumental to qualitative computational inquiries of fish schooling. Finally, we iterate that self-propelled agents are distinct from agents that are towed with a certain velocity. In agent based models, the latter are usually employed, but these do not correspond to self-propelled animals responding to non-linear interactions with the flow field.

The basis of this low-order model is depicted in \fig{fig:model}{}. In a system of $N$ dipole swimmers, a dipole located at $\mathbf{x}_{n}$ ($n = 1, 2, ..., N$) is decomposed into two vortices located at $\mathbf{x}_{n}^{\text{l}}$ and $\mathbf{x}_{n}^{\text{r}}$ with circulation strengths $\Gamma_{n}^{\text{l}}$ and $\Gamma_{n}^{\text{r}}$, separated by a constant distance $\ell$. The dipole swimmer travels with a bearing defined by $\alpha_{n}$ as depicted in the \fig{fig:model}{c}. Each finite dipole swimmer can change its vortex strengths as a means of controlling its bearing and speed. Following \cite{Tchieu:2012}, the equations of motion of $N$ self-propelled interacting finite dipoles are modified to allow each dipole to change its individual bearing and speed while simultaneously affecting the flow. Note that this model is different from the one used in \citep{Chate:2008} such that the generated flow field affects the bearing of swimmers and that the swimmers directly change the flow field when performing actions. We also note that the value $\ell$ can be related to a characteristic width of the swimmer D by matching the far-field dipolar strength of a body moving in an inviscid fluid to that of a finite dipole, resulting in $\ell= D/(2\sqrt{2 \pi})$. To proceed, a point $\mathbf{x}$ is mapped to the complex $z$--plane such that $\mathbf{x} = (x, y) \mapsto z = x + \mathrm i y$, where $\ii = \sqrt{-1}$. Therefore, given the position of a dipole $\mathbf x_{n} \mapsto z_{n}$, its two vortices of strengths $\Gamma_{n}^{\text{l}}$ and $\Gamma_{n}^{\text{r}}$, separated by a constant distance~$\ell$, are located at (\fig{fig:model}{})
\begin{equation}
	z_{n}^{\text{l}} = z_{n} + \frac{\ii \ell \eee{\ii \alpha_n}}{2} \hspace{0.5cm} \text{and} \hspace{0.5cm}
	z_{n}^{\text{r}} = z_{n} - \frac{\ii \ell \eee{\ii \alpha_n}}{2},\label{eq:mapComplex}
\end{equation}
respectively. The equations of motion that govern the system of $N$ finite dipoles derived from \citep{Tchieu:2012} are modified to read
\begin{subequations}
\label{eq:governingEom}
\begin{eqnarray}
	\dot{\conj{z}}_{n} &=& \frac{\left( \Gamma_{n}^{\text{l}} + \Gamma_{n}^{\text{r}} \right) \eee{	-\ii \alpha_{n}}}{4 \pi \ell} + \frac{w_{n}^{\mathrm{o}}(z_{n}^{\text{r}}) + w_{n}^{\mathrm{o}}(z_{n}^{\text{r}})}{2}, \label{eq:eomPosition}\\
	\dot{\alpha}_{n} &=& \frac{\Gamma_{n}^{\text{l}} - \Gamma_{n}^{\text{r}}}{2 \pi \ell^{2}} + \frac{\real{\left( w_{n}^{\mathrm{o}}(z_{n}^{\text{r}}) - w_{n}^{\mathrm{o}}(z_{n}^{\text{l}}) \right) \eee{\ii \alpha_{n}}}}{\ell},  \label{eq:eomAngle}
\end{eqnarray}
\label{eq:eom}
\end{subequations}
where Re$[\cdot]$ selects the real part of complex expressions and
\begin{equation}
	w_{n}^{\mathrm{o}}(z) = \sum_{j \neq n}^{N} \frac{1}{2 \pi \ii} \left( \frac{\Gamma_{j}^{\mathrm{l}}}{z -z_{j}^{\mathrm{l}}}  - \frac{\Gamma_{j}^{\mathrm{r}}}{z - z_{j}^{\mathrm{r}}} \right),
	\label{eq:desingularizedVelocity}
\end{equation}
is the interaction term due to all other dipoles in the environment. We emphasize that the first terms of Eq.~\eqref{eq:eomPosition} and Eq.~\eqref{eq:eomAngle} correspond to the dipole self-induced velocity and bearing rate when no other swimmers or background flow is present.  Therefore, these terms relate to the ability of an agent to affect its own speed and bearing.

We wish to stress the fact that the dipole model allows us to evolve the system in time by actually solving the Euler equations for an incompressible, inviscid flow and therefore the presence of a liquid environment is not simplistically modeled through ad hoc local interaction rules. The major advantage is that this formulation provides a neat distinction between social and hydrodynamic effects, unlike previous modeling approaches.

\subsection{Swimming gaits and maneuvering through circulation change}
\label{sec:actionsDipoles}
\begin{figure}
	\begin{center}
		\includegraphics[width=\textwidth]{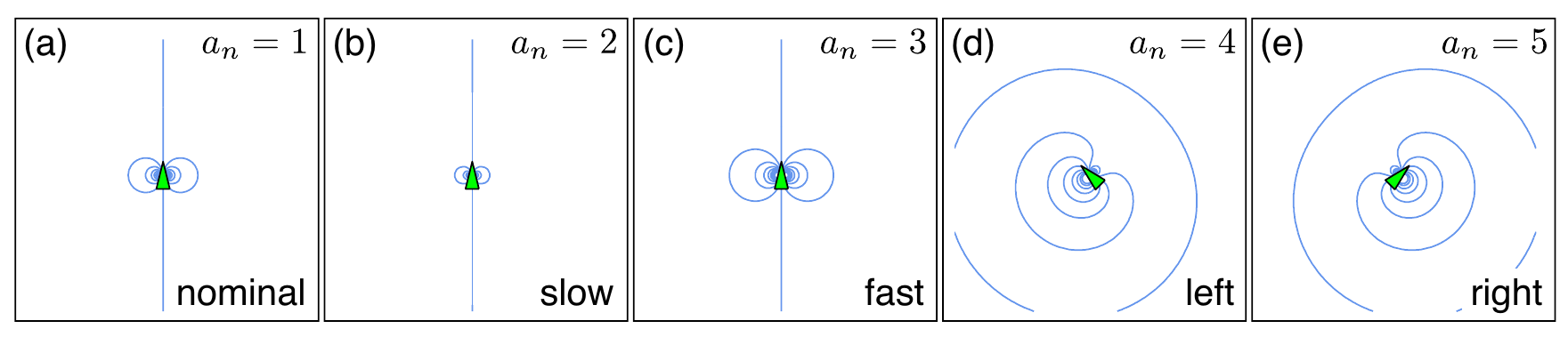}
		\caption{The set of actions available to each dipole swimmer (represented by triangles of width $D = 2 \ell \sqrt{2 \pi}$ pointing in the direction of travel). These actions are: (a) traveling straight at its nominal speed $v^{0}$, (b) traveling straight at a slower speed $v^{-}$, (c) traveling straight at a faster speed $v^{+}$, (d) making a left, and (e) making a right at a specified turn radius $\rho_{\mathrm{T}}$.  In the figures we also show the streamlines demonstrating how a swimmer affects the background flow field in the absence of all other dipole swimmers. The actions are mapped to the integer set $a_{n} = \{1, 2, 3, 4, 5\}$, respectively.}
		\label{fig:actions}
	\end{center}
\end{figure}

We extend the original finite dipole model to equip each swimmer with a set of gaits or actions, as depicted in \fig{fig:actions}{}. It can travel forward at three distinct speeds $v^0$ (nominal speed), $v^-= v^{0} - v^{\mathrm{A}}$ (slow), or $v^+=v^{0} + v^{\mathrm{A}}$ (fast), or turn left or right with turn radius $\rho^{\mathrm{T}}$ while traveling at speed $v^{0}$. These actions are realized by allowing each dipole swimmer to instantaneously adjust its vortex circulations $\Gamma_{n}^{\text{l}}$ and $\Gamma_{n}^{\text{r}}$. The five actions, mapped to integer values, adjust the circulation according to the rule
\begin{equation}
	\text{if} \quad  
 	\left\{
	\begin{aligned}
		\displaystyle a_{n} &= 1 \text{ : travel straight at $v^{0}$}  \quad & \longrightarrow & \quad  \Gamma_{n}^{\text{l}} = \Gamma_{n}^{\text{r}} = \Gamma^{0},  \\
		\displaystyle  a_{n}& = 2 \text{ : travel at straight $v^{-}$} \quad & \longrightarrow & \quad \Gamma_{n}^{\text{l}} = \Gamma_{n}^{\text{r}}= \Gamma^{0} - \Gamma^{\text{A}},  \\
		\displaystyle  a_{n} &= 3 \text{ : travel at straight $v^{+}$} \quad & \longrightarrow & \quad \Gamma_{n}^{\text{l}} = \Gamma_{n}^{\text{r}}=\Gamma^{0} + \Gamma^{\text{A}}, \\ 
		\displaystyle  a_{n} & = 4 \text{ : turn left with radius $\rho^{\mathrm T}$} \hspace{0.2cm} \quad & \longrightarrow & \quad \Gamma_{n}^{\text{l}} = \Gamma^{0} + \Gamma^{\text{T}}, \   \Gamma_{n}^{\text{r}} = \Gamma^{0} - \Gamma^{\text{T}} , \\
		\displaystyle  a_{n} &= 5 \text{ : turn right with radius $\rho^{\mathrm T}$} \quad & \longrightarrow & \quad \Gamma_{n}^{\text{l}} = \Gamma^{0} - \Gamma^{\text{T}},  \ \Gamma_{n}^{\text{r}} = \Gamma^{0} + \Gamma^{\text{T}},
		\label{eq:actionsExplicit}
	\end{aligned}
	\right.
\end{equation}
where $\Gamma^{0}, \Gamma^{\text{A}}, \Gamma^{\text{T}} > 0$. The nominal vortex strength $\Gamma^{0}$ is related to the cruise velocity $v^{0}$ and characteristic size $\ell$ such that $\Gamma^{0} = 2 \pi \ell v^{0}$.

The additional circulations $\pm \Gamma^{\mathrm{A}}$ and $\pm \Gamma^{\mathrm{T}}$ due to traveling fast, slow, or turning right or left, respectively, are fixed by the swimming parameters $v^{0}$, $v^{\mathrm A}$, and $\rho^{\mathrm{T}}$. These values are related to the nominal circulation by
\begin{subequations}
\begin{align}
	&\Gamma^{\text{A}} = \left( \frac{v^{\text{A}}}{v^{0}} \right)  \Gamma^{0},	\\ 
	&\Gamma^{\text{T}} = \left( \frac{\rho^{\text{T}}}{\sqrt{2 \pi} \ell} \right)^{-1} \Gamma^{0}.
\end{align}
\end{subequations}
The change in circulation in turn modifies the flow field and thus influences all swimmers in the system. Note that these actions are exclusive, i.e. a swimmer can only select one action at a time.

We emphasize that the use of five actions is a simplification with respect to naturally occurring swimmers, which are characterized by a large number of kinematic degrees of freedom and therefore have the ability to fine tune their gaits in response to environmental cues. However, a small number of actions drastically reduces the computational costs associated with identifying an optimal behavioral policy through reinforcement learning. Hence our choice to equip the agents with a limited repertoire of gaits.

\subsection{The Aoki-Couzin behavioral model with hydrodynamics interactions}
\label{sec:aokicouzin}
We examine how the classical  Aoki-Couzin model \citep{Aoki:1982, Couzin:2002} with \textit{a priori} defined interaction rules would perform in the presence of hydrodynamics. In particular, we considered the so called `dynamically parallel school' and `highly parallel school' behavioral rules, as detailed in \citep{Couzin:2002}.

In the Aoki-Couzin model collective behavior emerges  due to three \textit{a priori} specified rules among agents. Each agent tries to avoid collision with neighbors, aligns to the moving direction of the agents contained in a larger neighborhood and, finally, tends to approach the agents of an even larger neighborhood. Given these three rules, each agent first computes its desired direction and subsequently turns in order to meet it \citep{Couzin:2002}. By varying the size of the interaction regions, qualitatively different behaviors can be observed. In Table 1, we summarize the radii characteristic of `dynamically parallel school' and `highly parallel school', as indicated in \citep{Couzin:2002} and used here.

In order to cast the Aoki-Couzin model into the present dipole framework, each dipole agent determines its desired direction $\alpha_{\text{desired}}$ by following the specifications of \citep{Couzin:2002} and then adjusts its vortex circulations as follows
\begin{equation}
	\begin{aligned}
		\dot{\alpha}_{\text{desired}} &= \frac{\alpha_{\text{desired}}-\alpha_0}{\tau} \\
		\Gamma_{\text{add}} &= \pi \ell^2 \left( \dot{\alpha}_{\text{desired}} - \dot{\alpha}_{0} \right) \\
		\Gamma^\text{l} &= \Gamma^\text{l} + \Gamma_{\text{add}} \\
		\Gamma^\text{r} &= \Gamma^\text{r} - \Gamma_{\text{add}}
	\end{aligned}
\end{equation}
where $\tau$ and $\alpha_{0}$, $\dot{\alpha}_{0}$ are, respectively, the simulations time step and the agent's current bearing and bearing rate. Computing the new $\Gamma^\text{r}$ and $\Gamma^\text{l}$ corresponds to adapting gaits to match exactly the desired bearings, assuming the absence of all other swimmers. Once the circulations of each dipole are determined, the system of $N$ swimmers is evolved accounting for hydrodynamic coupling via the governing Eqs.~(\ref{eq:mapComplex}--\ref{eq:desingularizedVelocity}).

The behaviour of this model in the presence of hydrodynamic interactions is examined in Section ~\ref{sec:schooling}.

\begin{table}[h!]
\label{tab:behaviors}
\begin{center}
\begin{tabular}{lccc}
	Behavior & 			Zone of repulsion & 	Zone of orientation & 	Zone of attraction \\
	\hline
	Highly parallel & 		1\rule{0pt}{12pt} & 					12 & 				15 \\
	Dynamically parallel & 	1 & 					6 & 					10 \\
\end{tabular} \\
{Tab. 1:} \small{Radii of zones of interaction relative to `dynamically parallel school' and `highly parallel school' behaviors as detailed in \citep{Couzin:2002}. Radii are normalized by $\ell$.}
\end{center}
\end{table}

\subsection{Reverse engineering of dynamic interaction rules via reinforcement learning}
\label{sec:rl}
Swimmers are modeled as finite width dipoles with varying circulation strength. Their presence and actions affect the flow field and, in turn, all other swimmers. Due to this highly non-linear coupling it is virtually impossible to handcraft local interaction rules that allow the dipoles to coherently swim in any predefined, finite size schooling arrangement. Therefore agent based models, with \textit{a priori} defined rules cannot help assess the hydrodynamic properties of different schooling configurations. In turn we employ a reverse engineering approach to obtain the interaction rules among swimmers. We specify for the agents the goal of maintaining a given geometric arrangement and employ a reinforcement learning technique to identify an {}{appropriate} interaction policy. This approach relies on four key components: the \textit{reward} that encodes the agent's goal; the \textit{state} that formalizes what the dipole can sense of the surrounding environment; the \textit{actions}, that is the repertoire of gaits at disposal of the swimmer (Section~\ref{sec:actionsDipoles}); and finally a learning strategy based on trial and error.

In this study we employ a particular RL technique, namely the one-step $Q$-learning algorithm \citep{Sutton:1998}. Beside its algorithmic simplicity, $Q$-learning has been proven to converge to an optimal behavioral policy for finite Markov decision processes \citep{Watkins:1992}. In this setting, the swimming agent explores the environment and its experience is represented by the tuple $\langle s_{n},a_{n},r_{n},s_{n}' \rangle$, where $s_{n}'$ is the next state given the action $a_{n}$ taken from the current state $s_{n}$, and $r_{n}$ is the corresponding reward. An agent estimates by trial and error the action-value function $Q_{n}(s_{n},a_{n})$, i.e. the expected long term reward for taking action $a_{n}$ given the state $s_{n}$ (a schematic of this approach is depicted in \fig{fig:approach}{}). The action-value function $Q_n$ can be understood as a table or a matrix in which for every state-action entry the corresponding expected reward (estimated through the reward history) is stored. This table is consulted by the agent whenever an action has to be taken, and it is continuously updated as the system evolves. Therefore, $Q_{n}$ encodes the swimmer \textit{adaptive} decision making intelligence and the corresponding behavior is determined by choosing from $Q_{n}$, with probability $1-\epsilon$, the best action $a_{n}$ such that $a_{n} \rightarrow \max_{a_{n}} Q_{n}(s_{n},a_{n})$ from the current state $s_{n}$ ($\epsilon$-greedy selection scheme). The $\epsilon$--probability of choosing a non-optimal action allows the agent to explore new state-action $(s_{n},a_{n})$ pairs \citep{Sutton:1998}. Therefore, RL intrinsically accounts for noise through $\epsilon$, which can be related to the noise of natural schooling systems. Here, we use a shared policy approach among all swimmers to accelerate the learning process thus $Q_{n} = Q$ and all agents update $Q$ based on their personal experience. At every learning time interval $\delta t$, the swimmer updates the action-value function following $Q(s_n,a_n) = Q(s_n,a_n) + \varphi ( \Delta Q)_n$ for $n=1,\dots,N$ where $(\Delta Q)_{n} = r_n +\gamma\max_{a_{n}} Q(s_n',a_{n})-Q(s_n,a_n)$ and $0\le \varphi \le1$ is the learning rate and $0\le\gamma<1$ is the discount parameter which corresponds to the weight given to past experiences. We emphasize that learning individual policies, as opposed to the shared approach employed here, may allow agents finer behavioral tuning. For example, in the case of schooling, swimmers may adapt their policies depending on their location within the group. However, it has been empirically shown that the use of a shared policy reduces the time to convergence linearly with the number of agents \citep{Gazzola:2013a}. In our study this entails a hundreds-fold reduction in computational cost, hence the rational behind the choice of employing a shared approach. In the following the definitions of \textit{reward}, \textit{state} and \textit{action} are formalized.\\

\textit{Reward.} Since ultimately we are interested in investigating the hydrodynamic properties associated with different schooling geometries, we must first have the dipoles learn to swim in a given formation. This is achieved by setting that the goal of a swimmer is to follow a specified target point $\mathbf{x}_n^\mathrm{t}$ in a predefined arrangement as depicted in \fig{fig:model}{d}. This allows hundreds of self-propelled dipoles to learn how to swim coherently, a task out of reach for model based on handcrafted \textit{a priori} interaction rules (in Sections \ref{sec:optimalSchools} and \ref{sec:resultsOptimalSchools} we detail how optimal arrangements can be obtained). The swimmer's goal is mathematically cast into a numerical reward signal. The numerical reward is chosen to reflect how well the dipole swimmer can follow its assigned target point while doing the minimum amount of maneuvering so that $r_n = w_{d} \left(1-\frac{d_{n}}{\ell} \right) + w_{a}\cdot\eta_{a}$, where $\eta_{a}=0$ for traveling at $v_0$, $\eta_{a}=-1$ for accelerating and turning, and $\eta_{a}=1$ for decelerating. Weights are set to $w_{d}=0.9$ and $w_{a}=0.1$ thus $\max(r_{n}) = 1$. We note how the second term of $r_n$ penalizes swimmers that take unnecessary actions, while it favors those that reduce their effort by slowing down.\\

\textit{State.} Swimmers can sense their distance, $d_n = |\mathbf{x}_n^\mathrm{t} - \mathbf{x}_{n}|$, and orientation $\theta_n = \arg (\mathbf{x}_n^\mathrm{t} - \mathbf{x}_{n}) - \alpha_{n}$ with respect to their assigned target point $\mathbf{x}_n^\mathrm{t}$ within the school, as in \fig{fig:model}{d}. We wish to stress the fact that the dynamics of a swimmer is not mapped on a lattice. Swimmers are in fact free to move in the continuum two-dimensional space, while they adaptively adjust their gait in the attempt of maintaining their relative position within the school. The quantities $d_{n}$ and $\theta_{n}$ are each mapped into a set of $L = 30$ discrete states within the range $\Delta {d} = 10 D$ and $\Delta {\theta} = 2 \pi$ such that $s_{n} = \{ \min(L,\max(0,\lfloor d_{n} L/\Delta d \rfloor)), \min(L,\max(0,\lfloor \theta_{n} L/\Delta \theta \rfloor))\}$. In total, each swimmer has a state-space that consists of 900 states. The choice of a target point over the sensing of the neighbors  dramatically reduces the state space dimension (curse of dimensionality) allowing us to computationally tackle the problem. Moreover, it enables the study of structured schooling arrangements while it still captures the influence of the neighbors as they directly affect each other's dynamics through long range hydrodynamic interactions.\\

\textit{Actions.} Each dipole swimmer can perform five actions. It can travel forward at three distinct speeds $v^0$ (its nominal traveling speed), $v^+=1.1 v^{0}$, or $v^-=0.9 v^{0}$, or turn left or right with radius $r^T=10\ell$ while traveling at $v^{0}$. These actions are realized by allowing the dipole swimmer to adjust its vortex circulations $\Gamma_{n}^\text{l}$ and $\Gamma_{n}^\text{r}$ accordingly (Section \ref{sec:actionsDipoles}). Since for every state five different gaits are available, the overall state-action space amounts to 4500 entries, to be stored in a table representing the action-value function $Q(s_{n},a_{n})$.\\

An example further illustrating the working principles of RL is reported in Section \ref{sec:examplesimpleswimmer}.

\subsection{Optimization of schooling patterns via evolution strategies}
\label{sec:optimalSchools}
In the previous sections we established the algorithmic framework that allows self-propelled swimmers to learn how to keep their relative position within a school. We now look for optimal swimming configurations according to a desired metric (\textit{cost function}) using  the Covariance Matrix Adaptation Evolutionary Strategy (CMA-ES) \cite[]{Hansen:2003}. The CMA-ES has been proved effective in a number of fluid mechanics and biological problems, from the optimization of gait and morphology in swimmers \cite[]{Kern:2006a,Gazzola:2012,Rees:2013,Rees:2015} to the identification of virus traffic mechanisms \cite[]{Gazzola:2009}.

\begin{figure}
        \begin{center}
                \includegraphics[width=0.5\textwidth]{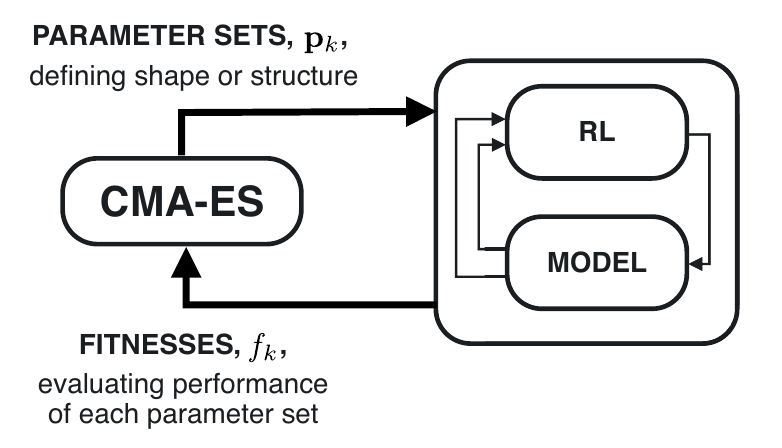}
                \caption{Schematic of CMA-ES optimizer coupled with RL and the low-order model. CMA-ES dispatches $k=1,...,p$ parameter sets $\mathbf{p}_{k}$ defining the school configuration. The RL framework allows agents to learn to swim in the formation characterized by $\mathbf{p}_{k}$. In return, CMA-ES receives a fitness $f_{k}$ that captures the effectiveness of collective swimming relative to each school $\mathbf{p}_{k}$.}
                \label{fig:cmaBlock}
        \end{center}
\end{figure}

The CMA-ES is a stochastic optimization algorithm that samples at each generation $p$ parameter vectors from a multivariate Gaussian distribution $\mathcal{N}$. Here each parameter vector encodes the geometric configuration of a schooling arrangement (\textit{Appendix}). The covariance matrix of the distribution $\mathcal{N}$ is then adapted based on successful past schools, chosen according to their corresponding cost function value $f$. In the present context, CMA-ES evolves schooling configurations based on a metric of swimming effectiveness. In order to evaluate the cost function, i.e. the performance, of each school geometry, the dipole swimmers must first learn through RL how to swim coherently in that specific arrangement. Then, after a learning period $\Delta T_{\text{learning}}$, the average cost function is evaluated by simulating the school for the time $\Delta T_{\text{eval}}$ (\textit{Appendix}). The value $f$ so computed is then returned to the optimizer, which uses it to select the best configurations and produce a new, more performant generation of school arrangements, until convergence to an optimal solution. This process is depicted in \fig{fig:cmaBlock}{}, while the definition of the cost function is given in the following.\\

\textit{Cost function.} Here the cost function implements a metric of swimming effort for the entire school to be minimized. We relate the effort of an individual swimmer compared to its nominal cruise effort, i.e. its additional circulation expenditure, by defining
\begin{equation}
\label{eq:individualEfficiency}
\Delta \Gamma_{n}^{\mathrm e}= \frac{1}{\Delta T_{\mathrm{eval}}} \int_{t}^{t+ \Delta T_{\mathrm{eval}}} ( \underbrace{|\Gamma_{n}^{\mathrm l} - \Gamma_{n}^{\mathrm r}|}_{\text{turning effort}} + \underbrace{ \Gamma_{n}^{\mathrm l} + \Gamma_{n}^{\mathrm r}}_{\text{forward effort}} - { 2 \Gamma^{0}}^{} ) \, \diffd t,
\end{equation}
where $\Gamma^{0}$ is the nominal strength of each vortex. Thus, the cost function for the entire collection of swimmers is defined as $f= \frac{1}{N \Gamma^{0}} \sum_{n=1}^N \Delta \Gamma_{n}^{\mathrm e}$. The change in circulation can be associated with the production of vorticity involved in accelerations or turning maneuvers of the swimmer, therefore to the swimming effort. We also note as reference that $f=0$ corresponds to cruise swimming of isolated dipoles.

\section{Results}
\label{sec:results}

\subsection{Learning process for an individual dipole swimmer}
\label{sec:examplesimpleswimmer}
Despite the formalism, the working principles of RL are rather simple. We illustrate them here with the aid of a simple but representative problem, before proceeding further.

We consider a dipole whose goal is to follow a prescribed trajectory. The trajectory is represented as a target point $\mathbf{x}^\mathrm{t}$ that moves by alternating straight runs to random turns of fixed radius $\rho^{\mathrm{T}}$. The dipole is aware at all times of its own bearing $\alpha$ and position $\mathbf{x}$ as well as of the target position $\mathbf{x}^\mathrm{t}$. At regular intervals $\delta t$, the agent is faced with the problem of choosing whether to turn left or right, accelerate, slow down or keep straight in order to accurately follow the target point. The dipole has not been instructed how to act given its relative position to the target, i.e. no \textit{a priori} local rules are enforced. Instead, every time the swimmer estimates its relative distance $d$ and orientation $\theta$ with respect to $\mathbf{x}^\mathrm{t}$, as illustrated in \fig{fig:model}{d}. This is equivalent to determining the state $s$, i.e. the agent's current situation. Since the intelligence or behavior of the swimmer is encoded as a multi-dimensional table or matrix, the continuous values of $d$ and $\theta$ are discretized into a number of integer values, as described in Section~\ref{sec:rl}. Once the state matrix entry is determined, the agent can consult the expected rewards stored in the matrix that are associated with taking each of the five aforementioned actions $a$. These values $Q(s,a)$ are constantly updated by the dipole and represent its past experience, and initially they are all set to zero. At this point the agent choses with probability $1-\epsilon$ the best action, i.e. the one with the largest $Q(s,a)$ value, and after pursuing it, the new distances $(d',\theta')$ from the target are estimated, defining the new state $s'$. Moreover, based on $d'$ the reward $r$ is assigned to the dipole. The policy is then improved by discounting the old estimate of the expected reward $Q(s,a)$ and complementing it with the new information $r$ according to the update rule described in Section~\ref{sec:rl}. This process is indefinitely repeated until convergence to an optimal behavioral policy.

\begin{figure}[h!]
	\begin{center}
		\includegraphics[width=0.85\textwidth]{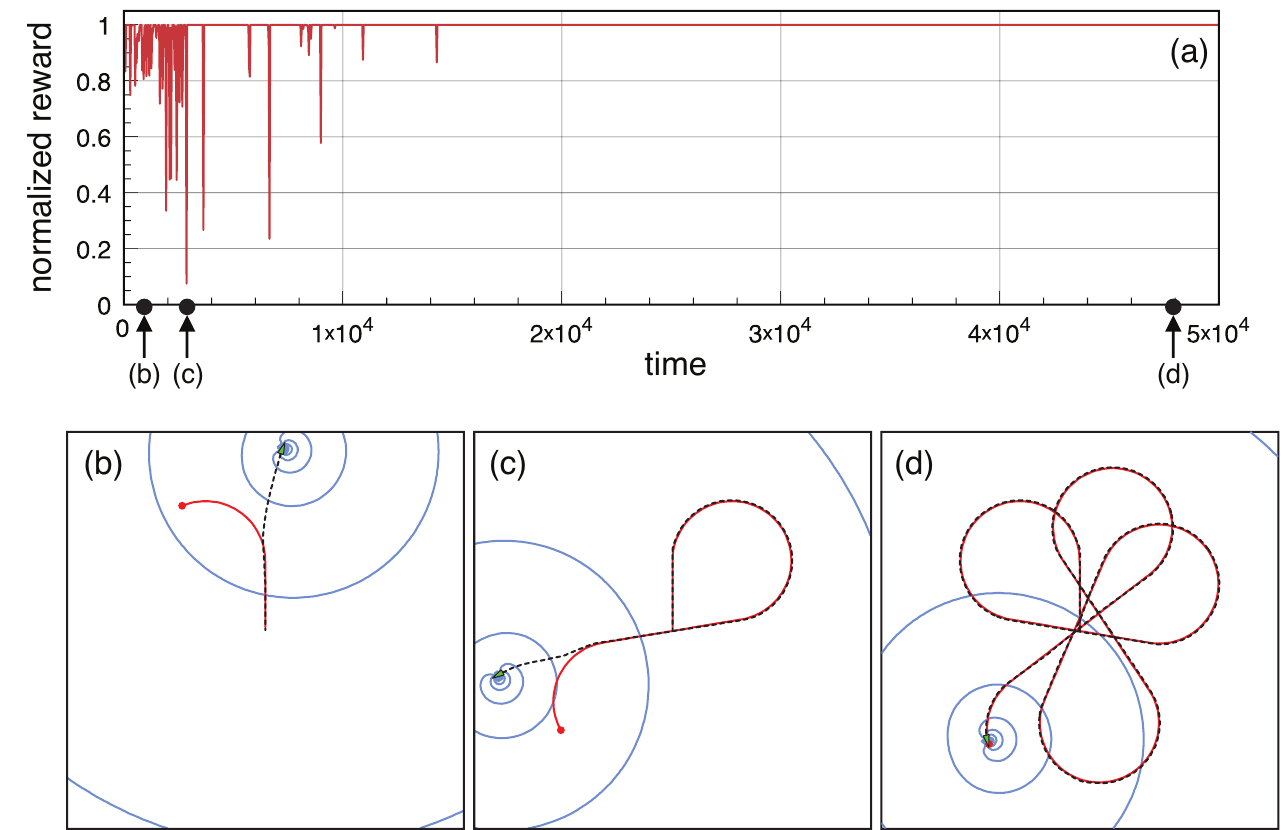}
		\caption{A single agent learns to optimally follow a moving target point $\mathbf{x}^\mathrm{t}$. Given in (a) is the time evolution of the agent's reward normalized by its maximum attainable value, based on the reward definition given in the main text. Panels (b), (c), and (d) correspond, respectively, to a non-adaptive, intermediate-adaptive, and well-adaptive learning stage. The starting time of panels (b--d) are marked on the $x$--axis of (a). Black dashed and solid red lines, correspond, respectively to the trajectories of the agent (green triangle) and the target point (red dot). Instantaneous streamlines (blue lines) are given as reference. Simulations are run in the domain $[0,1]\times[0,1]$ with $\ell = 3\times 10^{-3}$, $v^{0} = 5 \ell$, $v^{A} = 0.1 v^{0}$, $\rho^{\mathrm T} = 10 \ell$, $\delta t=0.1$, $\varphi=0.01$, $\gamma=0.98$, and $\epsilon=0.01$. Notation defined in Section \ref{sec:methods} and \textit{Appendix}.}
		\label{fig:singleDipole}
	\end{center}
\end{figure}

The evolution of the swimmer's reward over time is given in \fig{fig:singleDipole}{a} and examples of the agent trajectories during the learning process are given in \fig{fig:singleDipole}{b--d}. The process of improving the policy is seen in \fig{fig:singleDipole}{a}, where after an initial transient, the agent progressively learns how to maximize its numerical reward by accurately following $\mathbf{x}^\mathrm{t}$. In \fig{fig:singleDipole}{b}, the swimmer fails rather quickly as demonstrated when its path diverges from the target path at the first turn. Subsequently, in \fig{fig:singleDipole}{c} the swimmer follows adequately for a longer time before failing and in \fig{fig:singleDipole}{d}, the swimmer learns a policy that allows it to follow a pseudo-random path indefinitely.

\subsection{Classical agent based models versus learning agents in the presence of hydrodynamics}
\label{sec:schooling}
\begin{figure}
	\begin{center}
		\includegraphics[width=\textwidth]{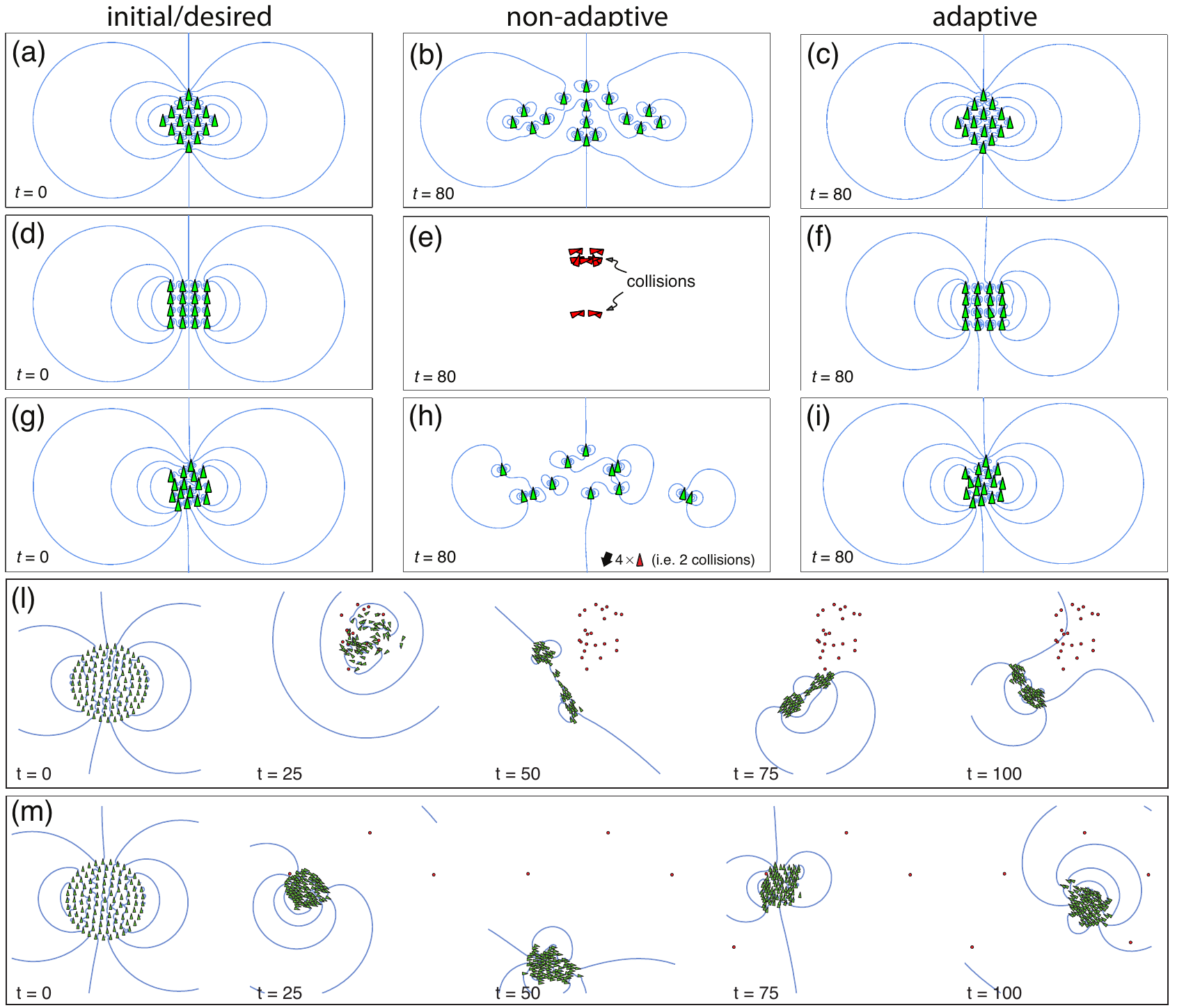}		
		\caption{(a,d,g) Initial/desired, (b,e,h) non-adaptive, and (c,f,i) adaptive (with policy learned from RL) swimming configurations for 16 dipole swimmers at specified times. Red dipole swimmers have experienced a collision (at which point they no longer move). Instantaneous streamlines (blue lines) are given as reference. Dipole swimmers are initialized on a diamond lattice in (a,b,c), a square lattice in (d,e,f) and randomly in a circular region of radius of $17.5 \ell$ in (g,h,i). In all initial configurations a minimum inter-dipole spacing of $10 \ell$ is enforced. (l,m) Time evolution of a school of $100$ agents obeying `dynamic parallel group' (l) and `highly parallel group' (m) models \citep{Couzin:2002} enhanced with hydrodynamic interactions. Simulations are run in the domain $[0,1]\times[0,1]$ with $\ell = 5\times 10^{-4}$, $v^{0} = 5 \ell$, $v^{A} = 0.1 v^{0}$, $\rho^{\mathrm T} = 10 \ell$, $\delta t=0.1$, $\varphi =0.01 $, $\gamma=0.98$, $\epsilon=0.01$. Notation defined in Section \ref{sec:methods} and \textit{Appendix}.}
		\label{fig:examples}
	\end{center}
\end{figure}

We first show that prescribed schooling patterns, including diamond and squares that have been proposed as favorable schooling patterns \citep{Weihs:1973}, are not robustly maintained without an adaptive dynamic response of the swimmers to the flow field. In \fig{fig:examples}{} we report the results of sixteen swimming agents attempting to school in several formations initialized ($t = 0$)  as shown in \figs{fig:examples}{a,d,g}. These initial patterns are characterized by a diamond-like, square-like and random arrangements.  With pre-specified forward swimming gait, the relative swimmer locations will result in varying hydrodynamic loads, thus implying a dynamic rearrangement of the swimmers.  Indeed, when the agents are assigned a specified swimming configuration, the simulated swimmers diverge from their relative positions and are prone to collisions with their neighbors due to flow-mediated interactions, consistently with \citep{Tchieu:2012}. As shown in \fig{fig:examples}{b} at $t = 80$, no collisions occur in the diamond-like configuration, but the swimmers are substantially strained apart. In \fig{fig:examples}{e}, the square arrangement causes all agents to collide, while random configurations lead to straining and collision effects simultaneously (\fig{fig:examples}{h}). A qualitative hydrodynamic explanation for the disruption of square-like and diamond-like formations in provided in Section \ref{sec:resultsOptimalSchools} and \fig{fig:dimondsquare}{}. In turn, in \figs{fig:examples}{c,f,i} we show that schooling patterns can be maintained by swimmers through an adaptive modification of their swimming gaits that accounts for hydrodynamic interactions using a reinforcement learning algorithm. Therefore, through RL the agents learn to adjust their swimming gaits to compensate for the varying hydrodynamic loads typical of a liquid environment. This can be related to the noisy environments and the corresponding response of swimmers in natural schooling systems.

We note that while the stability of diamond and square configurations has also been investigated by Tsang and Kanso \cite{Tsang:2013}, our approach fundamentally differs. In fact, Tsang and Kanso consider infinite, doubly periodic lattices of dipoles characterized by a single gait. The dipoles are then cleverly arranged so that the resulting flow field passively stabilizes the lattice, removing altogether the need of responding to varying loads by adapting swimming gaits, and the associated energetic costs. Here, we renounce to the assumption of infinite schools in favor of a more realistic description. As a consequence passive stabilization due to a stationary global flow field is no longer an option, hence the introduction of varying gaits and adaptive decision making. Therefore, our approach complements the results of \cite{Tsang:2013}, allowing us to study the hydrodynamic and energetic features associated with arbitrarily shaped, finite size schools.

\begin{figure}
	\begin{center}
		\includegraphics[width=\textwidth]{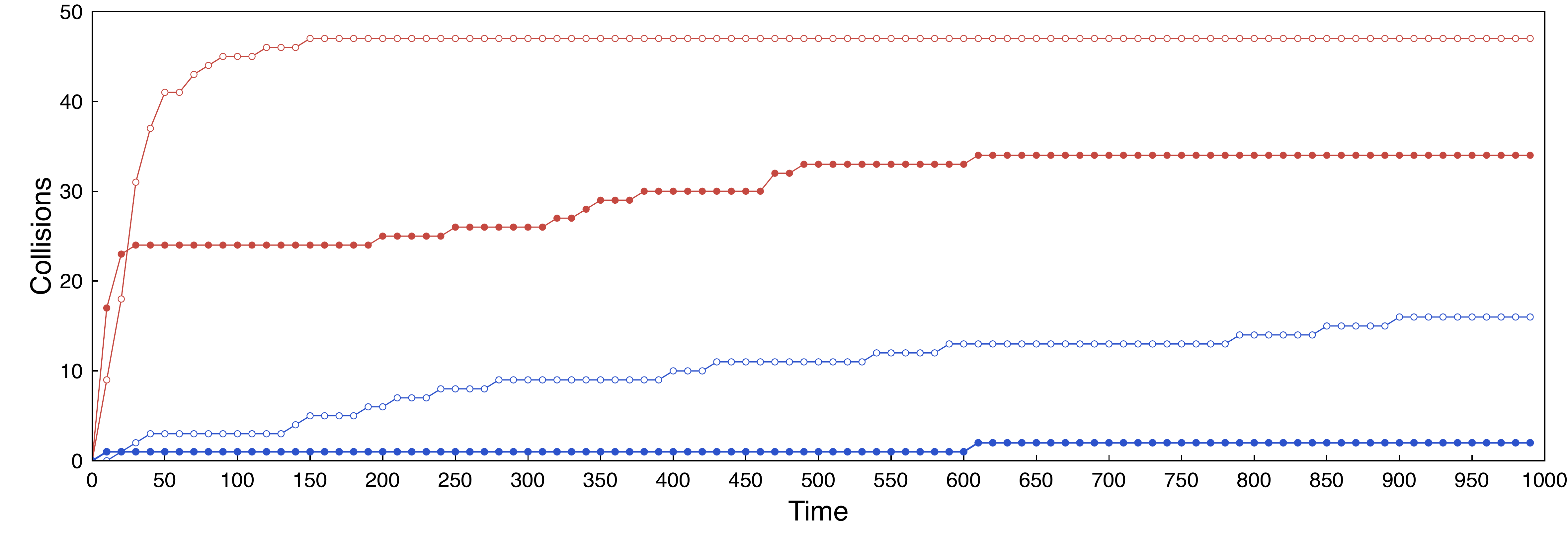}		
		\caption{We compare the quantitative response of the Aoki-Couzin model with and without hydrodynamics interactions in terms of number of agents' collisions in time. A collision is detected whenever the distance between two agents is found to be less than $2\ell$. Red and blue correspond, respectively, to `dynamically parallel group' and `highly parallel group' behavioral rules \citep{Couzin:2002}. Blank and solid circles refer, respectively, to simulations with and without hydrodynamics.}
		\label{fig:collisionsVs}
	\end{center}
\end{figure}

We also investigated the dynamics of the Aoki-Couzin behavioral model \citep{Aoki:1982, Couzin:2002} in the presence of hydrodynamic interactions (Section \ref{sec:aokicouzin}). This model relies on \textit{a priori} handcrafted local interaction rules among agents and does not explicitly account for the flow environment during the decision making process. The time evolution of a school of $100$ agents obeying `dynamic parallel group' and `highly parallel group' behavioral rules \citep{Couzin:2002} is shown in \fig{fig:examples}{l,m}. We find that the swimmers experience substantial straining and collisions, due to the hydrodynamic coupling. This behavior is a drastic departure from schooling patterns observed when employing the original models \citep{Couzin:2002}. Indeed, the number of collisions increases by $~40\%$ and $~700\%$ in, respectively, the `dynamic parallel group' and `highly parallel group' model in the presence of hydrodynamics (\fig{fig:collisionsVs}{}). These findings emphasize the role of the environment, especially in an hydrodynamic setting in which all agents are doubly connected through the flow. The fact that the non-linear response of the hydrodynamic system cannot be anticipated renders the definition of interaction rules by hand cumbersome, tedious, and ultimately not robust.

The agent based models with \textit{a priori} specified rules, such as the ones considered herein, are characterized by a large parameter space (size of each zone, attraction and repulsion weights, time step, etc.) and their results are known to be sensitive to these settings. In this study we have not explored the full parameter space  and as such it can not rule out the possibility that particular parameter combinations may allow dipole swimmers to maintain structured arrangements or to exhibit robust schooling dynamics. Nevertheless the present investigation raises two key issues. Firstly, the introduction of the flow environment modifies the dynamics associated with classical agent based model settings. This implies that to reproduce the behavior observed for a given instance of the Aoki-Couzin model, a new set of parameters has to be discovered. Since zone sizes, attractive and repulsive forces posses a well defined `social' meaning, the presence of the fluid affects these quantities and alters the nature of social interactions. Therefore, the characterization of social traits cannot prescind from accounting for the environment. Secondly, the sensitivity of classical models to parameter settings supports the need for rigorous, automatic procedures for the identification of local rules in the context of collective behavior. Indeed, we were unable to handcraft or derive through a direct optimisation process, any parameter set that enabled dipole swimmers to maintain a structured schooling formation. However this was readily achieved through the RL framework. We iterate that this finding does not represent a mathematical proof that handcrafted \textit{a priori} models do not lead robustly to schooling, but it strongly emphasizes the need for computational methods that robustly guide the systematic exploration of their parameter space.

\subsection{Optimal internal structure of a school of dipole swimmers}
\begin{figure}
	\begin{center}
		\includegraphics[width=\textwidth]{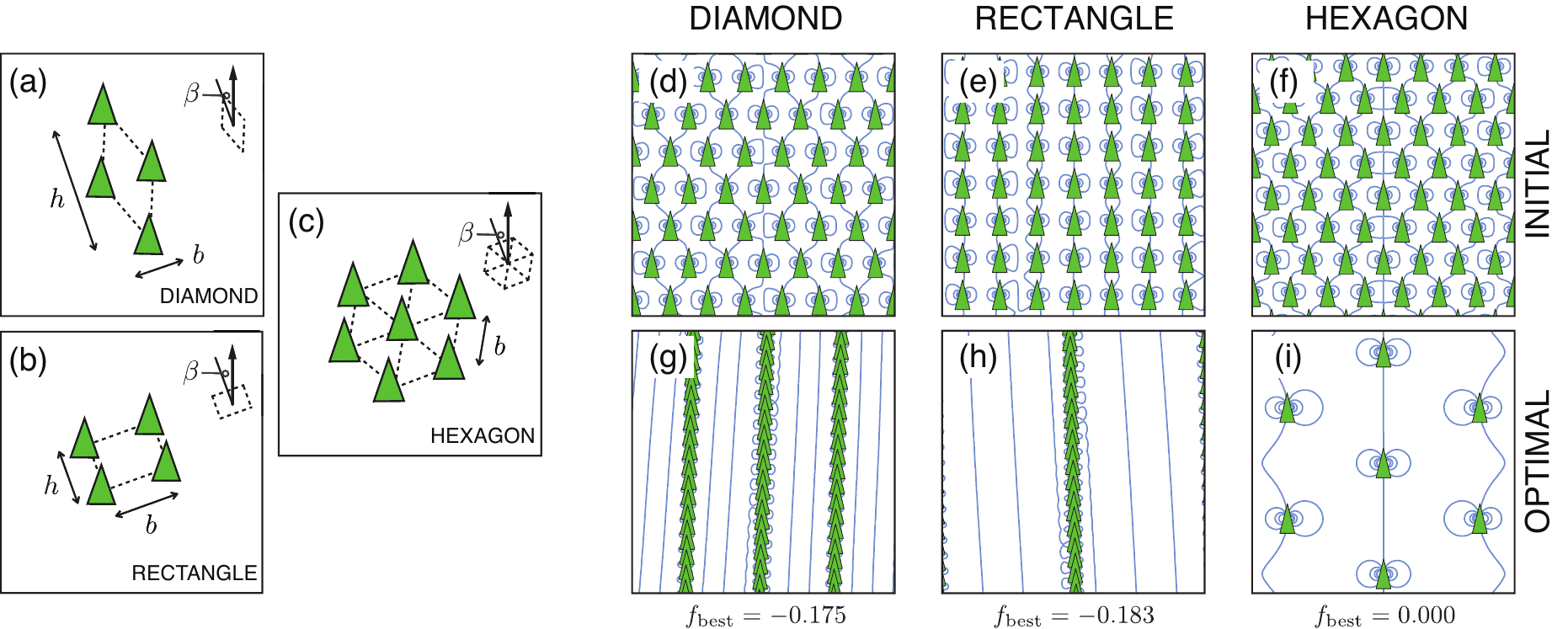}		
		\caption{The three different parameterizations for the internal structure of the school. We choose to investigate (a) diamond configurations parameterized by $\mathbf p = \{ b, h, \beta \}$, (b) rectangular configurations parameterized by $\mathbf p = \{ b, h, \beta \}$, and (c) hexagon configurations parameterized by $\mathbf p = \{ b, \beta \}$. In (c), all agents are equidistant to its nearest neighbors. The angle $\beta$ represents the difference in direction of travel with respect to the axis defining the structure. The initial guesses (d--f) where the fitnesses $f \approx 0$ and best optimized solutions (g--i) for the parameters defining the internal structure as defined according to (a--c). Population size is $p = 100$. Streamlines are given in blue. Optimal parameters are (g) $\mathbf p_{\text{best}} = \{ 49.70 \ell, 2.55 \ell, 0.02 \}$, (h) $\mathbf p_{\text{best}} =\{ 45.90 \ell, 5.20 \ell, -0.03 \} $, and (i) $\mathbf p_{\text{best}} = \{37.25 \ell, 0.52\}$ corresponding to the fitnesses $f_{\mathrm{best}} = -0.175, -0.184,$ and $0.000$. Swimmers tend to form striated patterns in diamond and rectangular configurations and in the case where dipole swimmers are required to be equidistant from one another, there is no apparent hydrodynamic benefit from staying in the collective. Simulations are run in a $[0,1]\times[0,1]$ box with $\ell = 5\times 10^{-4}$ $(\ell = 1\times 10^{-4}), v^{0} = 5 \ell$, $\rho^{\mathrm T} = 10 \ell$, $\delta t=0.1$, $\varphi =0.01 $, $\gamma=0.98$, $\epsilon=0.01$. Notation defined in Section \ref{sec:methods} and \textit{Appendix}. We note that the model relies on the assumption that swimmers are separated more than one characteristic length $\ell$ from one another \citep{Tchieu:2012}. This condition is met here with a minimum distance of $2.55\ell$ for case (g), corresponding to the spacing $b$ in our parameterization. This implies that swimmers could pack even tighter, but this scenario is found to be suboptimal.}
		\label{fig:structureStreams}
	\end{center}
\end{figure}

Since diamond lattices embedded in an infinite school have been suggested to be energetically favorable \citep{Weihs:1973,Tsang:2013}, we first characterize in terms of collective effort only the internal structure of schooling formations, disregarding edge effects related to the finite size of the school. The swimming effort of an individual is quantified by the variation of its dipole strength from a nominal value ($\Delta \Gamma_{n}^{\mathrm e}$, Section \ref{sec:optimalSchools}). This effort changes during swimming, according to the reinforcement learning algorithm, to overcome hydrodynamic noise. We therefore optimize the bulk school structure to minimise the cost function $f$ defined as the linear sum of the efforts of all the swimmers in the group (Section \ref{sec:optimalSchools}), with $f=0$ corresponding to cruise swimming of isolated swimmers. We consider three different parameterizations for generalized internal structure: (a) diamond (b) rectangle, and (c) hexagon configurations as depicted in \fig{fig:structureStreams}{a--c}. The parameter $\beta$ defines the angle between the swimming direction and axis defining the lattice structure (see the inset of \figs{fig:structureStreams}{a--c}). The hexagon formations are a subset of diamond formations, with the added restriction of swimmers to be equidistant to its nearest neighbors. To minimize edge effects, we generate a circular shape and fill it with a lattice of $N \approx 200$ agents. Collective effort is only evaluated for the interior agents and our criteria chooses for the 50 agents closest to the center of the school.

The starting configuration are given in \fig{fig:structureStreams}{d--f}, while the corresponding optimal solutions are reported in \fig{fig:structureStreams}{g--i}. The swimmers form striated patterns and get closely packed to one another in their traveling direction while separating as far as possible (given the bounds of the optimization search space) in the orthogonal direction. The packing in the direction of travel is limited by the capacity of the agents to stay in formation due to the strong flow-mediated interactions.

For the hexagon case (equidistant to its neighbors), from a starting configuration that is slightly detrimental for the school ($f_{1} = 0.001$), the optimizer finds a configuration that gives no added benefit to the collective, as the fitness of case \fig{fig:structureStreams}{i} is $f_{\mathrm{best}} \approx 0$. It is concluded that the constraint of equidistant swimmers does not allow the agents to pack in the direction of travel and thus is not favorable for reducing circulation expenditure.

The results reported in this section may be put in context in light of recent experiments that systematically investigated the thrust, power and efficiency performance of two side-by-side \citep{Dewey:2014} and in-line pitching airfoils \citep{Boschitsch:2014}. In the side-by-side case \citep{Dewey:2014}, it is found that the performance of individual airfoils are always anti-correlated except for perfectly in-phase or out-of-phase actuation (which may not be realistic or robust in a schooling system). This arrangement entails an overall constant system efficiency and the generation of a net torque, which needs to be compensated for in stable schooling arrays. In the context of our study,  these results suggest no foreseeable benefits associated with parallel swimming dipoles. Indeed, the identified striated patterns tend to minimize anti-correlation effects, by stretching lateral spacings as much as possible, effectively decoupling parallel dipoles. The in-line pitching airfoils case is more complex. For small spacings $s/\ell<0.5$ (where $s$ in the linear coordinate in the direction of travel and $\ell$ is the airfoil chord) the performance of the leading and trailing airfoils are anti-correlated. For larger spacings instead, only the trailing body is affected and the vortex shedding from the leading  airfoil has a prominent role in the observed dynamics. Extrapolating to multiple airfoils we may expect small spacings to be suboptimal. In fact, due to the strong anti-correlation every airfoil would experience both enhancing and disruptive effects. A larger spacing instead would allow, under the appropriate phase lag, all airfoils to benefit from flow coupling. The dipole model does not account explicitly for vortex shedding. As a consequence there is no phase lag or cutoff separation distance that controls the interactions of the leading and trailing dipoles. Nevertheless, the optimization process discards configurations characterized by small spacing since the strong anti-correlated dipole interaction poses control and learning problems and increases the swimming effort. Dipoles settle for larger spacing ($s/\ell\ge2.55$ in \fig{fig:structureStreams}{}, $s/\ell\ge1.35$ in \fig{fig:schoolshape}{}) to weaken anti-correlation effects, allowing for better stability and reduced effort, consistent with the above observations. We conclude that our findings, within the limits of our modeling approach, qualitatively capture the salient features associated with two-body swimming interactions \citep{Boschitsch:2014,Dewey:2014}.

\subsection{Optimal shape of a school of dipole swimmers}
\label{sec:resultsOptimalSchools}
We proceed by optimizing for collective swimming effort $f$ of the school configuration as a whole, where both internal structure and edge effects compete simultaneously. We initialise a school formation by arranging $N=100$ swimmers within a prescribed shape so as to maximize the distance from each other as well as the shape boundary (\textit{Appendix}). The schooling arrangement is specified by four parameters $\mathbf{p}=\{k_1, k_2, \phi_1, d_{\text{avg}}\}$, where $k_1$, $k_2$ and $\phi_1$ characterize its shape, while $d_{\text{avg}}$ determines its area $A=N\cdot d_{\text{avg}}^2$. The parameters corresponding to optimal schooling arrangement are identified through stochastic optimization (Section \ref{sec:optimalSchools}) that minimises $f$, with $f=0$ corresponding to  cruise swimming of isolated swimmers.

\begin{figure}
	\begin{center}
		\includegraphics[width=\textwidth]{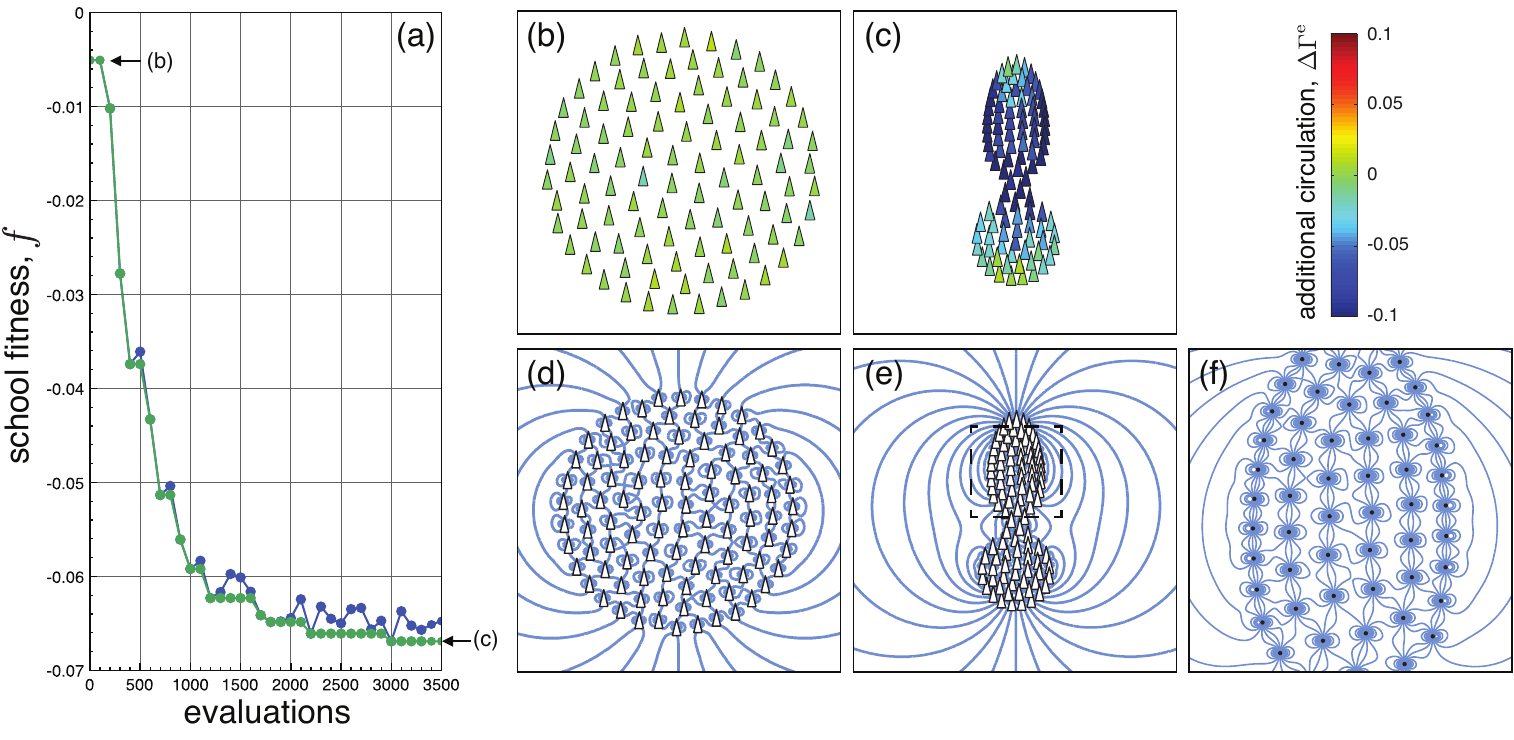}		
		\caption{(a) Evolution of cost function $f$ versus number of cost function evaluations. Population size per generation is $p = 100$. Blue and green lines correspond to, respectively, best solution in the current generation and best solution ever. Also given are school formations corresponding, respectively, to the (b) starting search point ($\mathbf{p}_1=\{3,1,1,1.57\}$, $f_1=-0.005$) and (c) best point ($\mathbf{p}_{\text{best}}=\{1.35,  0.5,  1.98, 2.42\}$, $f_{\text{best}}=-0.067$). The additional circulation change is computed for each swimmer and colored accordingly. Simulations are run in the domain $[0,1]\times[0,1]$ with $\ell = 5\times 10^{-4}$, $v^{0} = 5 \ell$, $\rho^{\mathrm T} = 10 \ell$, $\delta t=0.1$, $\varphi =0.01 $, $\gamma=0.98$, $\epsilon=0.01$.  All configurations correspond to a simulation time $t=100$. Notation defined in Section \ref{sec:methods} and \textit{Appendix}. Streamlines of the (g) starting search point, and (e) the best solution configuration. Given in (f) is a detailed view of the dashed box of (e) with the dipoles swimmers represented as black dots. As shown in (f), it is advantageous for the dipole swimmers to line up in striated patterns to aid in creating a favorable flow in the direction of travel. Indeed, the higher streamlines density in this direction is the footprint of a stronger longitudinal velocity field. We note that the dipole model relies on the assumption that swimmers are separated more than one characteristic length $\ell$ from one another \citep{Tchieu:2012}. This condition is met here with a minimum distance of $1.35\ell$ (first entry of the parameter vector $\mathbf{p}_{\text{best}}$). This implies that swimmers could pack even tighter, but this scenario is found to be suboptimal.} 
		\label{fig:schoolshape}
	\end{center}
\end{figure}

The course of the stochastic optimization is shown in \fig{fig:schoolshape}{a}. The schooling arrangement evolves from the initial circular shape of \fig{fig:schoolshape}{b} to the optimal `hourglass' solution of \fig{fig:schoolshape}{c}, reminiscent of shapes observed in nature for medium size schools \citep{Misund:1995,Parrish:2002}. The `hourglass' shape is associated with an $\sim80\%$ area contraction and with a 10-fold reduction in collective effort. The color coding in \fig{fig:schoolshape}{b,c} signifies the average swimming effort required by the dipoles to maintain their position in the school and illustrates how the circulation strength decreases for all swimmers, due to the more favorable arrangement.

The streamlines of the collective flow fields are illustrated in {\fig{fig:schoolshape}{d--f}. The swimmers in the `hourglass' school are found to align in striated patterns (\fig{fig:schoolshape}{f}), consistent with the findings of \fig{fig:structureStreams}{}, and synchronize to induce a stronger global dipolar field (\fig{fig:schoolshape}{d,e}), which is associated with a higher streamline density in the direction of travel and implies a stronger forward velocity. This formation allows swimmers to maintain their forward speed with a reduced individual effort  (smaller circulation strength). Furthermore, dipoles in the center of the necking region (\fig{fig:schoolshape}{c,e}) get the benefit of drafting from the swimmers in the front while at the same time they are pushed from the ones in the rear. In summary, packing to a smaller area while elongating the school shape, favors  striated swimming patterns (\fig{fig:schoolshape}{c,f}) so that swimmers benefit from flow-mediated interactions. We note that the optimal swimming pattern exhibits a swimming effort that outperforms the one of the Aoki-Couzin models by $~190\%$ ($f_{\text{a priori models}}\simeq0.06$).

Moreover, specified square-like and diamond-like formations are also shown to be detrimental in terms of collective and individual effort (\fig{fig:dimondsquare}{}), nevertheless their analysis is revealing of the hydrodynamic mechanisms at play in the `hourglass' optimal solution. Indeed, although the fitness for the entire school is $f=0.004$ for both \fig{fig:dimondsquare}{a} and \fig{fig:dimondsquare}{b}, swimmers exposed on the left and right edges of the diamond formation suffer from high circulation expenditures, while those in the square one in general do not. In the square formation, swimmers line up near the left and right edges of the school to help create greater net flow in the direction of travel. Conversely, swimmers in the front or the rear of the diamond school benefit from schooling, while those in the square formation do not due to the counter flow produced from their immediate left and right neighbors. The optimal school shape solution of \fig{fig:dimondsquare}{c} takes advantage of these two effects. The `hourglass' shape allows agents to line up near the boundary of the school and elongates in the traveling direction to reduce the counter flow from agents on the right and left edges.

\begin{figure}
	\centering
	\includegraphics[width=\textwidth]{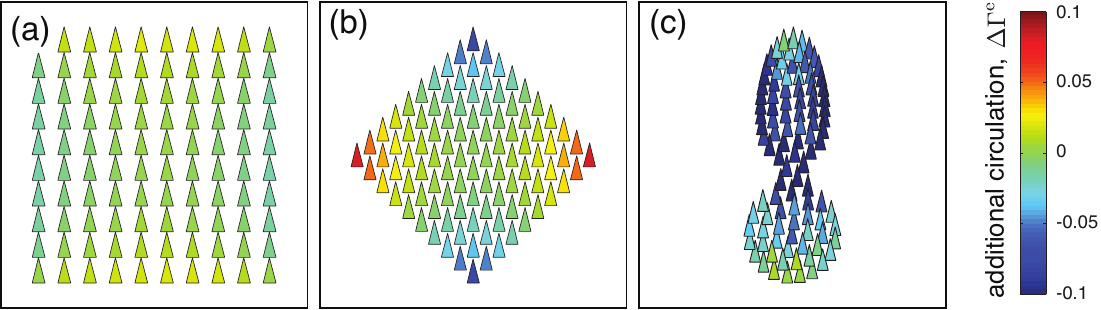} 
	\caption{Comparison between (a) square formation ($f = 0.004$), (b) diamond formation ($f = 0.004$) and (c) the optimal solution  ($f_{\text{best}}=-0.067$). The additional circulation change is computed for each swimmer and colored accordingly. Simulations are run in the domain $[0,1]\times[0,1]$ with $\ell = 5\times 10^{-4}$, $v^{0} = 5 \ell$, $\rho^{\mathrm T} = 10 \ell$, $\delta t=0.1$, $\varphi =0.01 $, $\gamma=0.98$, $\epsilon=0.01$. Notation defined in Section \ref{sec:methods} and \textit{Appendix}.}
	\label{fig:dimondsquare}
\end{figure}

\section{Discussion}
\label{sec:discussion}
We present a novel approach for studying schooling by coupling reinforcement learning and stochastic optimisation with swimming agents \citep{Tchieu:2012} that explicitly account for hydrodynamic interactions. Swimmers are modeled as self-propelled, finite width dipoles and have the capability to adjust their speed and orientation to cope with varying hydrodynamic loads. The dipoles represent the far field vortex wake of self-propelled natural swimmers, so that agents impart long range velocity fields on all other agents via their dipolar strengths. These non-linear hydrodynamic interactions critically affect schooling dynamics and the individual swimming patterns. We show that classical agent models, which rely on \textit{a priori} specified social rules, are not robust in the presence of hydrodynamics.
In order to compensate for hydrodynamics and to allow for schooling in the present agent based model, we reverse engineer the rules that are followed by the swimmers. This reverse engineering is achieved through a reinforcement learning algorithm, that creates mappings between the dynamic environment of the agents and their actions so as to maximize a numerical reward. Our approach differs from the widely used handcrafted \textit{a priori} behavioral rules, and allows us to examine how hydrodynamics affects swimmers' decision-making in schooling. We find that adaptive swimming policies are crucial for maintaining schooling formations.

We evaluate the effectiveness of various formation patterns through an evolution strategy algorithm that identifies optimal schooling shapes and swimmer arrangements. We find that schools exhibiting minimal collective effort are `hourglass' shaped and elongated in the swimming direction. Elongated shapes allow for drafting and pushing of swimmers arranged in internal striated patterns. Such internal striated patterns are found to be optimal independently of the overall shape of the school, qualitatively consistent with experimental observations \citep{Misund:1995,Parrish:2002,Boschitsch:2014,Dewey:2014}.

Moreover, it is found that a tight packing of swimmers inside the school allows them to exploit flow-mediated interactions in terms of collective effort by enhancing the global dipolar field. At the same time there is a limit  to the amount of packing, as it becomes increasingly difficult for the swimmers to stay in formation, due to stronger flow-mediated effects and increased probability of collisions. Such flow-mediated interactions can help explaining how certain fish travel in dense, elongated packs when migrating and foraging.

We wish to note that the present reverse engineering approach for the automatic identification of interaction rules relative to a goal can be readily generalized to other forms of collective behaviors, from car traffic to social aggregations. In the context of schooling, future work is concerned with extending the use of present learning and optimization techniques to two- and three-dimensional viscous flows of multiple swimmers at intermediate Reynolds numbers \citep{Gazzola:2014}. Reverse engineering techniques, as the ones proposed herein, can then be used to identify the various evolutionary traits that may have led to fish schooling.

\section{Appendix}

\subsection{Learning optimal behavior in a fluid mediated environment}

\subsubsection{Time integration and handling of collisions}
The set of ODEs given in Eqs.~(\ref{eq:eomPosition}-\ref{eq:eomAngle}) in the main text is numerically integrated using the forward Euler method so as to maintain flexibility with the decision selection process of the reinforcement learning algorithm. Since the equations become stiff as swimmers approach one another, the time step $\diffd t$ is computed according to the minimum distance $d_{\text{min}}$ between all dipole swimmers present in the environment. We bound $\diffd t$ with $\diffd t = \diffd t_{\text{max}}$ if $d_{\text{min}} \geq 2\ell$ and $\diffd t = \diffd t_{\text{min}} = 5e^{-4}$ if $d_{\text{min}} \leq \ell/2$.
We impose $\diffd t_{\text{max}}$ to be dependent on the nominal velocity $v^0$ such that $\diffd t_{\text{max}} = (5\ell C)/v^0$. For the simulations given here, $C = 0.005$.

Dipoles tend to collide if they are in close proximity with another dipole \citep{Tchieu:2012}. In a collision of two dipoles, each dipole stops and annihilates their respective circulation strengths. We manage this phenomenon by labeling colliding dipoles as `dead' if $\diffd t = \diffd t_{\text{min}}$. The dipoles' circulations are artificially set to zero, hence, dead dipoles no longer influence the other swimmers in the flow.

\subsubsection{Error analysis of non-adaptive and adaptive schools of dipole swimmers}
As noticed in \fig{fig:examples}{} of the main text and in \fig{fig:errors}{} here, in the presence of hydrodynamic interaction a given schooling arrangement is not maintained robustly without an adaptive dynamic response to the flow. We quantitatively demonstrate that schooling patterns can be maintained by swimmers through reinforcement learning. This is illustrated in \fig{fig:errors}{} in which the errors (average distance to the target points $e = \sum_{n}^{N} d_{n}/\ell$) between the simulations of \fig{fig:examples}{} are compared.

\begin{figure}
	\begin{center}
		\includegraphics[width=0.6\textwidth]{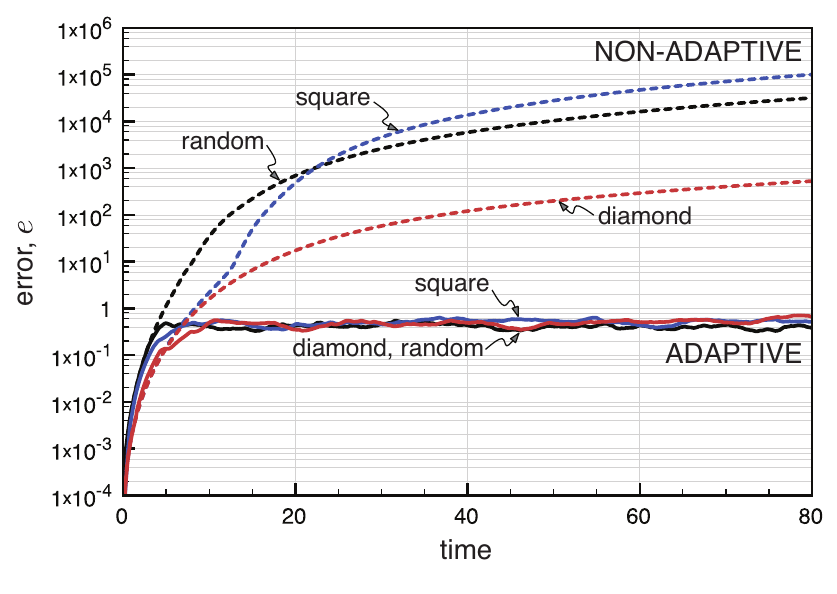}		
		\caption{The time evolution of the sum of all errors between all agents and their respective assigned target points. The non-adaptive school simulations (dotted lines) of diamond (red), square (blue), and random (black) formations correspond to \fig{fig:examples}{b,e,h} of the main text, respectively. The error in adaptive school simulations (bottom set of solid lines) of diamond (red), square (blue), and random (black) formations correspond to \fig{fig:examples}{c,f,i} of the main text, respectively. The error is defined as $e = \sum_{n}^{N} d_{n}/\ell$, i.e. the average distance to the target points.}
		\label{fig:errors}
	\end{center}
\end{figure}

\subsection{Optimal schooling formations}
We seek optimal schooling configurations by combining RL with an evolutionary strategy. The algorithm of choice is the Covariance Matrix Adaptation Evolutionary Strategy (CMA-ES) in its multi-host, rank-$\mu$ and weighted recombination form \cite{Hansen:2003,Gazzola:2011}. The robustness of CMA-ES is mainly controlled by the population size $p$ \cite[]{Hansen:2003}. In this work, as a tradeoff between robustness and fast convergence, we set $p=100$ for all optimization campaigns. Bounds of the search space are enforced during the sampling through a rejection algorithm.

In this strategy CMA-ES determines the optimal \textit{configuration} based on the metric of swimming effectiveness $f$, while RL determines the optimal \textit{policy} for an agent to follow its target point under any configuration requested by CMA-ES. Therefore, every cost function evaluation entails CMA-ES dispatching a parameter set defining the geometry of the school, then a RL training period ($\Delta T_{\text{training}}=10000$) that allows the dipoles to learn how to swim in the given arrangement, followed by an evaluation interval ($\Delta T_{\text{eval}}=100$) in which the school effectiveness is measured and returned to CMA-ES.

\subsubsection{CMA-ES settings and parameterization for the optimization of school shape}
We evaluate how the overall school shape affects collective effort and optimize its effectiveness in terms of Eq.~\eqref{eq:individualEfficiency} of the main text. To do so, we create a school by designing a general external shape and arranging the swimmers inside the boundary by placing them maximally distant from one another and the boundary. The shape of the school dictated by a cubic spline-based parameterization introduced in~\citep{Rossinelli:2011b}. According to this parameterization, the external school shape reads as $c = S\left( \phi \right)$, where $c$ is the radial distance from the origin in the polar plane, $\phi \in \left[ 0,\pi \right]$ is the angle, and $S$ is the piecewise polynomial of the cubic spline. The spline control points (red dots in \fig{fig:spline}{}), expressed in polar coordinates, are $\left(c_0,0\right)$, $\left(c_1,\phi_{1}\right)$ and $\left(c_2,0\right)$ with the radii defined as $c_1 = k_1 \cdot c_0$ and $c_2 = k_2 \cdot b_0$, $k_1$ and $k_2$ being constants. The school shape is completed by mirroring the obtained spline profile. Unlike~\citep{Rossinelli:2011b}, the area of the school shape $A(S)$, and therefore $c_0$, is controlled through an extra parameter, namely the average distance $d_{\text{avg}}$ between swimmers, such that $b_0:A(S)=N\cdot d_{\text{avg}}^2$, where $N$ is the number of dipoles. Therefore, the shape of the school relies on the four parameters $\mathbf{p} = \{d_{\text{avg}}$, $k_1$, $k_2$, $\phi_1 \}$. These free parameters are varied within the search space $[1,5]\times[0.1,2]\times[0.1,2]\times[\pi/6,5\pi/6]\in\mathbb{R}^4$ during the optimization.
\begin{figure}
	\centering
	\includegraphics[width=0.8\textwidth]{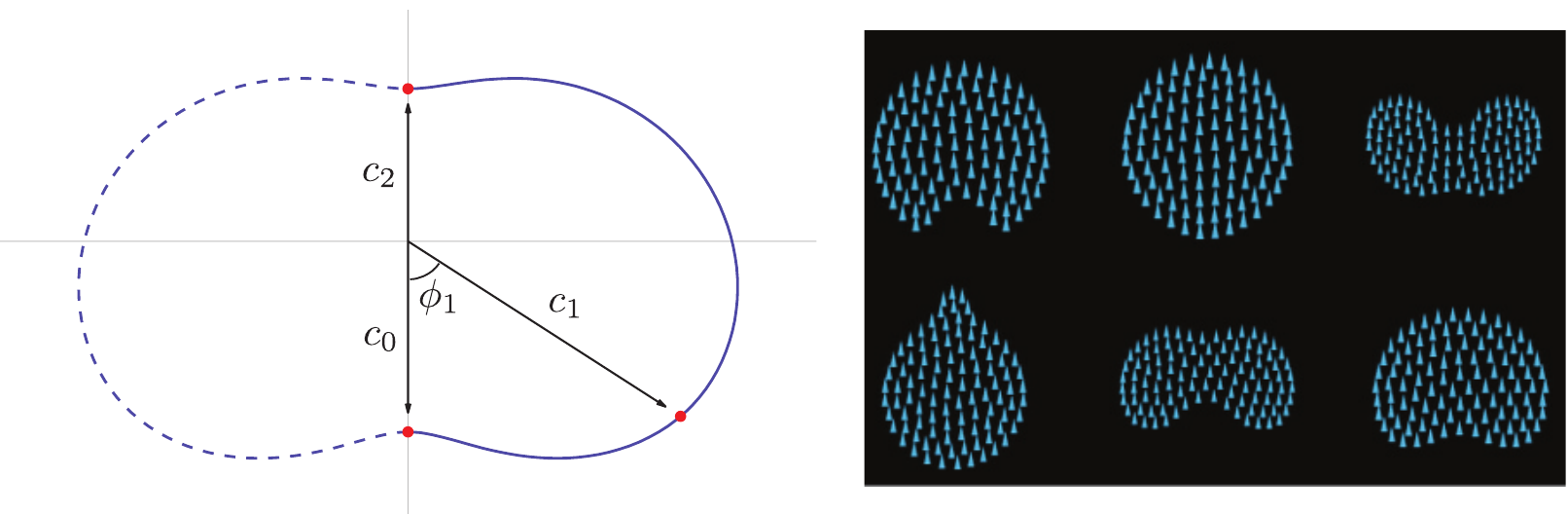} 
	\caption{Parametrization of the shape of the school and a representative subset of candidate solutions. The parameterization creates the solid line (blue) and is mirrored to create the symmetric complement (dashed line). The control points are depicted as red dots.}
	\label{fig:spline}
\end{figure}

\subsubsection{CMA-ES settings for the optimization of internal lattice structure of a school}
The free parameters $h, b,$ and $\beta$ for the diamond, rectangular, and hexagonal search space (see \fig{fig:structureStreams}{a--c} of the main text) range from $[0,50 \ell]\times[0,50 \ell]\times[0,\pi/2]\in\mathbb{R}^3$, $[0,50 \ell]\times[0,50 \ell]\times[0,\pi/2]\in\mathbb{R}^3$, and $[0,50 \ell]\times[0,\pi/6]\in\mathbb{R}^2$, respectively, during the optimization.

\bibliographystyle{unsrtnat}

\end{document}